\documentclass[preprint,aps,prd]{revtex4}
\newcommand{\bea}{\begin{eqnarray}}
\newcommand{\eea}{\end{eqnarray}}
\newcommand{\beq}{\begin{equation}}
\newcommand{\eeq}{\end{equation}}
\newcommand{\nn}{\nonumber}
\def\k{{\vec k}}
\def\x{{\vec x}}
\def\/{\over}

\begin{document}

\title{Spontaneous excitation of an accelerated atom \\in a spacetime with a reflecting
plane boundary}
\author{Hongwei Yu}
\affiliation{ CCAST(World Lab.), P. O. Box 8730, Beijing, 100080,
P. R. China and Department of Physics and Institute of  Physics,\\
Hunan Normal University, Changsha, Hunan 410081,
China\footnote{Mailing address}}

\author{Shizhuan Lu }
\affiliation{Department of Physics and Institute of  Physics,\\
Hunan Normal University, Changsha, Hunan 410081, China}

\begin{abstract}
We study a two-level atom in interaction with a real massless
scalar quantum field in a spacetime with a reflecting boundary.
The presence of the boundary modifies the quantum fluctuations of
the scalar field, which in turn modifies the radiative properties
of atoms. We calculate
 the rate of change of the mean atomic energy
of the atom for both inertial motion and uniform acceleration. It
is found that the modifications induced by the presence of a
boundary make the spontaneous radiation rate of an excited
inertial atom to oscillate near the boundary and this oscillatory
behavior may offer a possible opportunity for experimental tests
for geometrical (boundary) effects in flat spacetime. While for
accelerated atoms, the transitions from ground states to excited
states are found to be possible even in vacuum due to changes in
the vacuum fluctuations induced by both the presence of the
boundary and the acceleration of atoms, and this can be regarded
as an actual physical process underlying the Unruh effect.
\end{abstract}
\maketitle

\baselineskip=16pt

\section{Introduction}

Spontaneous emission is one of the most important features of
atoms and so far mechanisms such as vacuum fluctuations
\cite{Welton48, CPP83}, radiation reaction \cite{Ackerhalt73}, or
a combination of them \cite{Milonni88} have been put forward to
explain why spontaneous emission occurs. The ambiguity in physical
interpretation arises from different choices of ordering of
commuting operators of atom and field in a Heisenberg picture
approach to the problem. Significant progress has been made by
Dalibard, Dupont-Roc and CohenTannoudji(DDC), who argued in
Ref.\cite{Dalibard82} and Ref.\cite{Dalibard84} that there exists
a symmetric operator ordering that the distinct contributions of
vacuum fluctuations and radiation reaction to the rate of  change
of an atomic observable are separately Hermitian. If one demands
such a ordering, each contribution can possess an independent
physical meaning. The DDC prescription resolves the problem of
stability for ground-state atoms when only radiation reaction is
considered and the problem of ``spontaneous absorption" of atoms
when only vacuum fluctuations are taken into account. Using this
prescription one can show that for ground-state atoms, the
contributions of vacuum fluctuations and radiation reaction to the
rate of change of the mean excitation energy cancel exactly and
this cancellation forbids any transitions from the ground state
and thus ensures atom's stability. While for any initial excited
state,
 the rate of change of atomic energy acquires equal
contributions from vacuum fluctuations and from radiation
reaction.

Recently,  Audretsch, M\"ueller and Holzmann
\cite{Audretsch94,1Audretsch95,2Audretsch95} have generalized the
formalism of DDC \cite{Dalibard84} to evaluate vacuum fluctuations
and radiation reaction contributions to the spontaneous excitation
rate and radiative energy shifts of an accelerated two-level atom
interacting with a scalar field in a unbounded Minkowski space. In
particular, their results show that when an atom is accelerated,
then the delicate balance between vacuum fluctuations and
radiation reaction is altered since the contribution of vacuum
fluctuations to the rate of change of the mean excitation energy
is modified while that of the radiation reaction remains the same.
Thus transitions to excited states for ground-state atoms become
possible even in vacuum. This result not only is consistent with
the Unruh effect \cite{Unruh}, but also provides a physically
appealing interpretation of it. The Unruh effect, which is closely
related to the Hawking radiation of black holes, predicts that a
linearly accelerated two-level particle detector becomes excited
when moving through the Minkowski vacuum and it behaves as if it
were immersed (inertial) in a bath of thermal radiation at the
Unruh temperature proportional to its acceleration, when coupling
with massless scalar fields is considered. However, it is worth
noting that the equality between the behavior of uniformly
accelerated two-level particle detectors coupled with massless
scalar fields and the inertial detectors lying at rest in a
thermal bath may not be valid in general when the massless scalar
field is replaced by other fields \cite{VM}. Let us illustrate
this as follows. Couple an Unruh-DeWitt detector (two-level
monopole) with energy gap $\Delta E$ to a massive scalar field
with mass $m>\Delta E$. The excitation per proper time of this
detector when it is uniformly accelerated with proper acceleration
$a = $constant) goes for $\Delta E << m$ as
\begin{equation}
 P/T \sim a \int_0^\infty dx x K_0^2 [\sqrt{x^2 +(m/a)^2}] \;,
 \label{PT}
\end{equation}
where $K_\nu(z)$ is a Bessel function of imaginary argument. This
is clearly non-vanishing because the external agent is making work
on the detector. On the other hand, the same detector lying
inertial(!) at rest in a thermal bath with temperature $T=a/2\pi$
is  unable to excite because $\Delta E < m \leq \omega$, where
$\omega$ is the energy of the massive scalar particles. Of course,
this fact does not challenge by any means the Unruh effect because
what the Unruh effect does state is that  Eq.~(\ref{PT}) can be
recovered by using Fulling's quantization in conjunction with the
fact that the Minkowski vacuum is a thermal state of Rindler
particles \cite{VM}.

In this sense, the Unruh effect is intrinsically related to the
effects of modified vacuum fluctuations induced by the
acceleration of the atom (or detector) in question.  On the other
hand, however, It is well-known that the presence of boundaries in
a flat spacetime also modifies the vacuum fluctuations of quantum
fields, and it has been demonstrated that this modification (or
changes) in vacuum fluctuations can lead to a lot of novel
effects, such as the Casimir effect \cite{Cas}, the light-cone
fluctuations when gravity is quantized \cite{Yu1,Yu2,YF,Yu3}, and
the Brownian (random) motion of test particles in electromagnetic
vacuum \cite{YF04,YC}, just to name a few. Therefore, it remains
interesting to see what happens to the radiation properties of
accelerated atoms found in Ref.\cite{Audretsch94} when the vacuum
fluctuations are further modified by the presence of boundaries.
In this paper, following the formalism developed by Audretsch and
M\"uller\cite{Audretsch94}, we will calculate the effects of
modified vacuum fluctuations and radiation reaction due to the
presence of a reflecting plane boundary upon the spontaneous
excitation of  both an inertial and a uniformly accelerated atom
interacting with a quantized real massless scalar field. Let us
note here that the response rate of a uniformly accelerated source
interacting with a massless real scalar field in the presence of
boundaries has recently been discussed \cite{AC}.

The paper is organized as follows, we will review the formalism
developed in Refs.~\cite{Audretsch94} in Sec. II, then apply it to
the case of an inertial atom in Sec. III and to the case of an
accelerated atom in Sec. IV. Finally we will conclude with some
discussions in Sec. V

\section{The general formalism for vacuum fluctuations and radiation reaction}

To study how the spontaneous emission of atoms  is modified by the
presence of a reflecting plane boundary, we examine a simple case:
a two-level atom in interaction with a real massless scalar
quantum field which obeys the Dirichlet boundary condition
$\phi(x)|_{z=0}=0$. Here we have assumed that the reflecting plane
boundary is located at $z=0$ in space.  Let us consider a
pointlike two-level atom on a stationary space-time trajectory
$x(\tau)$, where $\tau$ denotes the proper time on the trajectory.
The stationary trajectory guarantees the existence of stationary
atomic states, $|+ \rangle$ and $|- \rangle$, with energies
$\pm{1\/2}\omega_0$ and a level spacing $\omega_0$.  The atom's
Hamiltonian which controls the time evolution with respect to
$\tau$ is given, in Dicke's notation \cite{Dicke54},  by
 \beq H_A (\tau)
=\omega_0 R_3 (\tau),  \label{atom's Hamiltonian}
 \eeq
 where $R_3 =
{1\/2} |+ \rangle \langle + | - {1\/2}| - \rangle \langle - |$ is
the pseudospin operator commonly used in the description of
two-level atoms\cite{Dicke54}. The free Hamiltonian of the scalar
quantum field that governs its time evolution with respect to
$\tau$ is
 \beq
  H_F (\tau) = \int
d^3 k\, \omega_\k \,a^\dagger_\k a_\k
    {dt\/d \tau}. \label{free Hamiltonian}
 \eeq
Here $a^\dagger_\k$, $a_\k$ are the creation and annihilation
operators with momentum $\k$. Following Ref. \cite{Audretsch94},
we assume that the interaction between the atom and the quantum
field is described by a Hamiltonian
 \beq
 H_I (\tau) = \mu R_2 (\tau) \phi (
x(\tau)),
\label{interaction Hamiltonian}
 \eeq
 where $\mu$ is a
coupling constant which we assume to be small, $R_2 = {1\/2} i (
R_- - R_+)$, and $R_+ = |+ \rangle \langle - |$, $R_- = |- \rangle
\langle +|$. The coupling is effective only on the trajectory
$x(\tau)$ of the atom.

We can now write down the Heisenberg equations of motion for the
atom and field observables. The field is always considered to be
in its vacuum state $|0 \rangle$.  We will separately discuss the
two physical mechanisms that contribute to the rate of change of
atomic observables: the contribution of vacuum fluctuations and
that of radiation reaction. For this purpose, we can split the
solution of field $\phi$ of the Heisenberg equations  into two
parts:  a free or vacuum part $\phi^f$, which is present even in
the absence of coupling, and a source part $\phi^s$, which
represents the field generated by the interaction between the atom
and the field. Following DDC\cite{Dalibard82,Dalibard84}, we
choose a symmetric ordering between atom and field variables and
consider the effects of $\phi^f$ and $\phi^s$ separately in the
Heisenberg equations of an arbitrary atomic observable G. Then, we
obtain the individual contributions of vacuum fluctuations and
radiation reaction to the rate of change of G. Since we are
interested in the spontaneous emission of the atom,  we will
concentrate on the mean atomic excitation energy $\langle
H_A(\tau) \rangle$. The contributions of vacuum fluctuations(vf)
and radiation reaction(rr) to the rate of change of $\langle H_A
\rangle$ can be written as ( cf.
Ref.\cite{Dalibard82,Dalibard84,Audretsch94} )
 \bea
 \left\langle {d H_A (\tau) \/ d\tau}
\right\rangle_{vf} &=&
    2 i \mu^2  \int_{\tau_0}^\tau d \tau' \, C^F(x(\tau),x(\tau'))
    {d\/ d \tau} \chi^A(\tau,\tau'), \label{general form of vf}\\
\left\langle {d H_A (\tau) \/ d\tau} \right\rangle_{rr} &=&
    2 i \mu^2
    \int_{\tau_0}^\tau d \tau' \, \chi^F(x(\tau),x(\tau')) {d\/
d \tau}
    C^A(\tau,\tau'),
    \label{general form of rr}
 \eea
with $| \rangle = |a,0 \rangle$ representing the atom in the
state $|a\rangle$ and the field in the  vacuum state $|0
\rangle$.
Here the statistical
functions of the atom, $C^{A}(\tau,\tau')$ and $\chi^A(\tau,\tau')$, are
defined as
\bea
C^{A}(\tau,\tau') &=& {1\/2} \langle
a| \{ R_2^f (\tau), R_2^f (\tau')\}
    | a \rangle,\label{general form of Ca} \\
\chi^A(\tau,\tau') &=& {1\/2} \langle a| [ R_2^f (\tau), R_2^f
(\tau')]
    | a \rangle \label{general form of Xa}
 \eea
 and those of the field are
 \bea
 C^{F}(x(\tau),x(\tau')) &=& {1\/2}{\langle} 0| \{ \phi^f
(x(\tau)), \phi^f(x(\tau')) \} | 0 \rangle,
\label{general form of
Cf}\\
\chi^F(x(\tau),x(\tau')) &=& {1\/2}{\langle} 0| [
\phi^f(x(\tau)),\phi^f (x(\tau'))] | 0 \rangle.
\label{general
form of Xf}
 \eea
$C^A$ is called the symmetric correlation function
of the atom in the state $|a\rangle$, $\chi^A$ its linear
susceptibility.  $C^F$ and $\chi^F$ are the Hadamard function and Pauli-Jordan or
 Schwinger function of the field respectively.

The explicit forms of the statistical functions of the atom are
given by \bea C^{A}(\tau,\tau')&=&{1\/2} \sum_b|\langle a | R_2^f
(0) | b
    \rangle |^2 \left( e^{i \omega_{ab}(\tau - \tau')} + e^{-i
\omega_{ab}
    (\tau - \tau')} \right), \label{explicit form of Ca}\\
\chi^A(\tau,\tau') & =& {1\/2}\sum_b |\langle a | R_2^f (0) | b
\rangle |^2
    \left(e^{i \omega_{ab}(\tau - \tau')} - e^{-i \omega_{ab}(\tau -
\tau')}
    \right), \label{explicit form of Xa}\eea
where $\omega_{ab}= \omega_a-\omega_b$ and the sum runs over a
complete set of atomic states. Using the method of images, at a
distance $z$ from the boundary, the statistical functions of the
field can be written as
 \bea
C^{F}(x(\tau),x(\tau'))&=&{1\/ 8\pi^2}\bigg\{-{1\/(\Delta t+ i
\epsilon)^2 - |\Delta \x|^2} -
    {1 \/(\Delta t- i \epsilon)^2-|\Delta \x|^2}\nonumber\\&&+{1 \/(\Delta t+
i
\epsilon)^2-[(x-x')^2+(y-y')^2+(z+z')^2]}\nonumber\\
&&+{1\/(\Delta t- i \epsilon)^2 -
[(x-x')^2+(y-y')^2+(z+z')^2]}\bigg\},\label{explicit form of Cf} \\
\chi^F(x(\tau),x(\tau'))&=&{ i\/ 4 \pi}\epsilon(\Delta
t)\{\delta({\Delta t}^2-((x-x')^2+(y-y')^2+(z+z')^2))\nonumber \\
&&-\delta({\Delta t}^2-|{\Delta \x}|^2)\}, \label{explicit form of
Xf}
    \eea
where $\Delta t=t(\tau) - t(\tau')$, $\Delta \x = \x (\tau) - \x
(\tau')$, and
\beq
\epsilon(\Delta t)=\cases{+1 \qquad for \qquad \Delta
t>0\cr
-1 \qquad for \qquad \Delta t<0 \cr} \label{sing function}
\eeq\\
is the sign function.

\section{Spontaneous emission from a uniformly moving atom}
\label{sec:inertial}
 In the present Section, we apply the formalism given in the proceeding Section to study the spontaneous emission
 of an inertial atom in the presence of a reflecting plane boundary.  For an inertial atom moving
in the $x$-direction with a constant velocity $v$ at a distance
$z$ from the plane, one has \beq t(\tau)  = \gamma \tau, \qquad
    x (\tau)  = x_0 +  v \gamma \tau,\qquad  y(\tau)=y_0, \qquad  {z}(\tau)=z
    \label{trajectory of atom}
    \eeq
where $\gamma = (1-v^2)^{-{1\/2}}$. From the general forms
Eq.(\ref{explicit form of Cf}) and Eq.(\ref{explicit form of Xf}),
we can easily obtain the statistical functions of the field
 \bea
C^F(x(\tau),x(\tau'))&=&-{1\/ 8 \pi^2} \bigg( {1\/(\tau-\tau'
    + i \epsilon)^2}-{1\/(\tau-\tau'
    + i \epsilon)^2-4z^2}\nonumber\\ &&+ {1\/(\tau-\tau' - i \epsilon)^2}
  -{1\/(\tau-\tau' - i \epsilon)^2-4z^2}
    \bigg),\label{from of Cf with trajectory}\\
\chi^F(x(\tau),x(\tau'))&=&- {i \/ 4\pi
}\epsilon(\gamma(\tau-\tau'))\{\delta((\tau-\tau')^2)-
\delta((\tau-\tau')^2-4z^2)\}.
    \label{from of Xf with trajectory}
 \eea

We can now evaluate  Eq.(\ref{general form of vf}) and
Eq.(\ref{general form of rr}) using the statistical functions
given above, With a substitution $u=\tau - \tau'$,  we get,  for
the contribution of the vacuum fluctuations to the rate of change
of atomic excitation energy,
 \bea
\left\langle {d H_A (\tau) \/ d\tau} \right\rangle_{vf} &=&
   {\mu^2 \/ 8 \pi^2} \sum_b \omega_{ab} |\langle a | R_2^f (0)
    |b\rangle |^2 \int_{-\infty}^{+\infty} du\,\bigg({1\/
    (u+ i \epsilon)^2}+{1\/ (u-i \epsilon)^2}\nonumber \\ &&-{1\/
    (u-2z+ i \epsilon)(u+2z+ i \epsilon)}-{1\/ (u-2z-i \epsilon)(u-2z-i
    \epsilon)}\bigg)e^{i\omega_{ab}u},
    \label{explicit form of vf}
    \eea
and for that of radiation reaction
 \bea
\left\langle {d H_A (\tau) \/ d\tau} \right\rangle_{rr} &=&
i{\mu^2 \/ 4 \pi}
 \sum_b \omega_{ab} |\langle a | R_2^f (0)
 |b\rangle |^2 \bigg\{\int_{-\infty}^{+\infty} du\,
 {\delta(u)\/u}e^{i \omega_{ab} u}\nonumber \\
 &&-\int_{0}^{\infty} du\,
 {\epsilon(\gamma u)\/4z}[\delta(u+2z)+\delta(u-2z)]
 e^{i \omega_{ab} u}\nonumber \\
 &&+\int_{-\infty}^{0} du\,
{\epsilon(-\gamma u)\/4z}[\delta(-u+2z)-\delta(-u-2z)]
 e^{i \omega_{ab} u}\bigg\},
 \label{explicit form of rr}
 \eea
where we have extended the range of integration to infinity for
sufficiently long times $\tau-\tau_0$. The integrals in
Eq~(\ref{explicit form of vf}) and Eq~(\ref{explicit form of rr})
can be evaluated via the residue theorem to get
 \bea
 \left\langle {d H_A (\tau) \/ d\tau}
\right\rangle_{vf} &=&
    -{\mu^2 \/2 \pi}  \sum_{\omega_a > \omega_b}
    |\langle a | R_2^f (0) | b \rangle
    |^2\left[{1\/2}\omega_{ab}^2-{\omega_{ab}\/4z}\sin(2z\omega_{ab})\right] \nonumber \\
  &&+{\mu^2 \/2 \pi}  \sum_{\omega_a <\omega_b}
    |\langle a | R_2^f (0) | b \rangle
    |^2\left[{1\/2}\omega_{ab}^2-{\omega_{ab}\/4z}\sin(2z\omega_{ab})\right],
     \label{result form of vf}
     \eea
     for the contribution
of vacuum fluctuation to the rate of change of the atomic
excitation energy and
 \bea \left\langle {d H_A (\tau) \/
d\tau} \right\rangle_{rr} &=&
    -{\mu^2 \/2 \pi}  \sum_{\omega_a > \omega_b}
    |\langle a | R_2^f (0) | b \rangle
    |^2\left[{1\/2}\omega_{ab}^2-{\omega_{ab}\/4z}\sin(2z\omega_{ab})\right]\nonumber \\
  &&-{\mu^2 \/2 \pi} \sum_{\omega_a <\omega_b}
    |\langle a | R_2^f (0) | b \rangle
    |^2\left[{1\/2}\omega_{ab}^2-{\omega_{ab}\/4z}\sin(2z\omega_{ab})\right].
    \label{result form of rr}
  \eea
for that of radiation reaction

 A few comments are now in order here.
One can see from Eq.~(\ref{result form of vf}) that for an atom
initially in the excited state ($|a \rangle=|+\rangle$), only the
first term ($\omega_a>\omega_b$) contributes
 and one has $\left\langle {d H_A (\tau) \/ d\tau}
\right\rangle_{vf}<0$. While for an atom initially in the ground
state ($|a \rangle=|-\rangle$), only the second term survives
($\omega_a>\omega_b$) and so $\left\langle {d H_A (\tau) \/ d\tau}
\right\rangle_{vf}>0$. Therefore, if only contributions of vacuum
fluctuations are considered, both spontaneous excitation and
de-excitation would equally occur. This leads to the well-known
problem of spontaneous absorption for a ground state atom in
vacuum. On the other hand, Eq.~(\ref{result form of rr}) shows
that radiation reaction always makes the atom to lose energy since
$\left\langle {d H_A (\tau) \/ d\tau} \right\rangle_{rr}<0$, no
matter if it is in the ground or excited state. This leads to a
problem similar to the instability of atoms in classical
electrodynamics. However, by adding the contributions of vacuum
fluctuations and radiation reaction, we obtain the total rate of
change of the atomic excitation energy
 \bea \left\langle {d H_A \/
d\tau} \right\rangle_{tot} &=&
    \left\langle {d H_A  \/ d\tau} \right\rangle_{vf}
    + \left\langle {d H_A \/ d\tau} \right\rangle_{rr}\nn\\
    &=& -{\mu^2
\/2 \pi} \left(\; \sum_{\omega_a
> \omega_b}
    |\langle a | R_2^f (0) | b \rangle
    |^2(\omega_{ab}^2-{\omega_{ab}\/2z}\sin(2z\omega_{ab}))\right). \label{explicit form of total change}
 \eea
 It follows that
 for an atom in the ground
state $(\omega_a < \omega_b)$, the effects of both contributions
exactly cancel, since each term in $\left\langle {d H_A (\tau) \/
d\tau} \right\rangle_{vf}$ is canceled exactly by the
corresponding term in $\left\langle {d H_A (\tau) \/ d\tau}
\right\rangle_{rr}$.
 Therefore, the presence of a plane boundary conspires to  modify  the vacuum fluctuations and radiation reaction
in such a way that the delicate balance  between the vacuum
fluctuations and radiation reaction found in
Ref.~\cite{Audretsch94} in absence of boundaries remains  and this
ensures the stability of ground-state inertial atoms in vacuum
with a reflecting boundary.

Eq.~(\ref{explicit form of total change}) gives the radiation rate
of an excited atom. The corrections induced by the presence of the
boundary are represented by $z$ dependent terms in all the above
results and they are oscillating functions of $z$ and
$\omega_{ab}$. Let us note first that corrections only change the
rate of change of atomic energy quantitatively but not
qualitatively since we always have
$\omega_{ab}^2-{\omega_{ab}\/2z}\sin(2z\omega_{ab})\geq 0$.
Secondly, as $z$, distance of the atom from the boundary,
approaches infinity, our results reduce to those of the unbounded
Minkowski space \cite{Audretsch94}
 as expected.
Thirdly, for a given atom,  the radiation rate is a function of
$z$ and it could either be enhanced or be weakened as compared
with the case without any boundary, depending on the atom's
distance to the plane boundary. Finally, the radiation rate
becomes zero, as the atom is placed closer and closer to the
boundary. This can be understood as a result of the fact that the
scalar field vanishes on the boundary and so does the interaction
Hamiltonian Eq.(\ref{interaction Hamiltonian}).

Let us now calculate the Einstein A coefficient for the
spontaneous emission of inertially moving atoms in the presence of
the boundary. For this purpose, following Ref.~\cite{Audretsch94},
we can obtain a differential equation for the atomic excitation
energy in order $\mu^2$, \beq \left\langle {d H_A (\tau) \/ d\tau}
\right\rangle_{tot} =
    - {\mu^2 \/8 \pi} \omega_0 \left( {1\/2} \omega_0 + \langle
H_A (\tau)
\rangle\right)\left(1-{\sin(2z\omega_0)\/2z\omega_0}\right).
\label{differential equation of total change}
 \eeq
 The solution of (\ref{differential
equation of total change}) is
\beq
\langle H_A (\tau) \rangle = -
{1\/2} \omega_0 + \left( \langle H_A (0)
    \rangle + {1\/2} \omega_0 \right) e^{-A \tau },
\label{solution of atomic energy}
 \eeq
 the familiar exponential decay to the atomic ground state $\langle H_A \rangle =
-{1\/2}\omega_0$. The spontaneous emission rate is given by the
Einstein A coefficient of the scalar theory:
\beq A  = {\mu^2 \/ 8
\pi} \omega_0\left(1-{\sin(2z\omega_0)\/2z\omega_0}\right).
\label{Einstein coefficient}
\eeq
We see once again that the rate of spontaneous emission is modified by the presence of the boundary and the Einstein
coefficient is function of $z$ for a given atom.

\section{Uniformly accelerated atom}

Let us now turn to the case in which the atom is uniformly
accelerated in a direction parallel  to the reflecting plane
boundary.  We assume that the atom is at a distance $z$ from the
boundary and is being accelerated in the $x$ direction with a
proper acceleration $a$. Specifically, the atom's trajectory is
described by
 \beq
 t(\tau)= {1\/a} \sinh a \tau,
\qquad x(\tau) = {1\/a} \cosh a \tau, \qquad z(\tau)=z, \qquad
y(\tau) = 0\;. \label{AccTraject}
 \eeq
 The
statistical functions of the field for the trajectory
Eq.~(\ref{AccTraject}) can be evaluated from their general forms
Eq.~(\ref{explicit form of Cf}) and Eq.~(\ref{explicit form of
Xf}). After some calculations, we obtain
 \bea
 C^F(x(\tau), x(\tau'))
&=& -{a^2 \/32\pi^2} \bigg( {1\/\sinh^2[{a\/2}
    (\tau-\tau') +i \epsilon]}+ {1\/\sinh^2[{a\/2} (\tau-\tau')
-i \epsilon]}\nn\\
    && \;-{1\/\sinh^2[{a\/2}
    (\tau-\tau') +i \epsilon]-(az)^2}- {1\/\sinh^2[{a\/2} (\tau-\tau')
-i \epsilon]-(az)^2}\;\biggr) \nn\\
 \label{AccVf}\\
\chi^F(x(\tau),x(\tau'))  &=& -\frac{i
}{8\pi}\frac{a}{\sinh\frac{a}{2}(\tau-\tau')} \,
    \bigg(\delta (\tau-\tau')\nonumber\\
&& -\frac{1}{2\sqrt{1+(az)^2}} \,
    \delta \bigl(\tau-\tau'-\frac{2}{a}\sinh^{-1}(az)\bigr)\nonumber\\
    &&+\frac{1}{2\sqrt{1+(az)^2}} \,
    \delta \bigl(\tau-\tau'+\frac{2}{a}\sinh^{-1}(az)\bigr)\biggr)\;.
 \label{AccRf}
 \eea
 With the help of the following integral, which can be readily
 evaluated by residues,
 \bea
 \int_{-\infty}^{\infty}&&\biggl(\frac{1}{\sinh^2(\frac{a}{2}
    u +i \epsilon)-(az)^2}+ \frac{1}{\sinh^2({a\/2} u
-i \epsilon)-(az)^2}\;\biggr)e^{i\omega_{ab}u}du\nonumber\\
&=&\left(1+\frac{2}{e^{2\pi|\omega_{ab}|/a}-1}\right)\frac{4\pi\sin
(\frac{2\omega_{ab} \sinh^{-1}(az)}{a})}{a^2z\sqrt{1+(az)^2}}\;,
 \eea
we can calculate the contribution of vacuum fluctuation to the
rate of change of the atomic excitation energy to get
 \bea
 \left\langle {d H_A (\tau) \/ d\tau} \right\rangle_{vf} &=&
    -{\mu^2\/ 2\pi} \Biggl[ \sum_{\omega_a > \omega_b}
\omega_{ab}^2
    |\langle a | R_2^f (0) | b \rangle |^2
    \;f(\omega_{ab},a,z)\;\left( {1\/2} + {1\/ e^{{2 \pi\/a} \omega_{ab}} -1} \right)
\nn\\
    && -\sum_{\omega_a < \omega_b} \omega_{ab}^2 |\langle a |
R_2^f (0) |
    b \rangle |^2 \;f(\omega_{ab},a,z)\;\left( {1\/2} + {1\/ e^{{2 \pi\/a}
|\omega_{ab}|} -1}
    \right)
\;\Biggr]\;,\nn\\
 \label{vfExcitation}
 \eea
where
 \bea
 f(\omega_{ab},a,z) = 1-\frac{1}{2\omega_{ab} z \sqrt{1+(az)^2}}\sin
\left(\frac{2\omega_{ab}z \sinh^{-1}(az)}{az}\right)\;.
 \eea
 Comparing the above result with Eq.~(56) of Ref.~\cite{Audretsch94}, one can see
 that the function $f(\omega_{ab},a,z)$ gives the modification
 induced by the presence of the boundary.
When $z\rightarrow \infty$, the function $f(\omega_{ab},a,z)$
approaches 1 and we recover the result obtained in
Ref.~\cite{Audretsch94} for a uniformly accelerated atom in a
unbounded Minkowski space as expected. On the other hand, if
$(az)\rightarrow 0$, one has
 \bea
f(\omega_{ab},a,z)\approx 1
-{\sin(2\omega_{ab}z)\/2\omega_{ab}z}\;. \label{appxf}
 \eea
 Let us note that  $(az)\rightarrow 0$ can be fulfilled
either by keeping $a$ at fixed finite value and letting $z$
approach zero or keeping $z$ fixed and letting $a$ go zero. For
the former case, $f(\omega_{ab},a,z)\approx 0$. This reveals that
as the atom gets closer and closer to the boundary the
contribution of the vacuum fluctuations to the rate of change of
the atomic excitation energy dies off in an oscillatory manner, no
matter if the atom is in inertial motion (refer to
Eq.~(\ref{result form of vf})) or is accelerated as long as the
proper accelerated is finite. While for the latter case, plugging
Eq.~(\ref{appxf}) into Eq.~(\ref{vfExcitation}), one recovers the
result for an inertially moving atom, i.e., Eq.~(\ref{result form
of vf}).

Similarly, one has for the contribution of radiation reaction,
 \bea
 \left\langle \frac{d H_A (\tau) }{ d\tau} \right\rangle_{rr} &=&
    -\frac{\mu^2}{2 \pi}  \sum_{\omega_a > \omega_b}
    |\langle a | R_2^f (0) | b \rangle
    |^2\biggl[\,\frac{1}{2}\omega_{ab}^2-\frac{\omega_{ab}}{4z\sqrt{1+(az)^2}}\nonumber \\
    &&\times\sin\biggl(\,\frac{2\omega_{ab}z \sinh^{-1}(az)}{az}\,\biggl)\,\biggr]
  -\frac{\mu^2}{2 \pi} \sum_{\omega_a <\omega_b}
    |\langle a | R_2^f (0) | b \rangle
    |^2 \nonumber \\
    &&\times\biggl[\,\frac{1}{2}\omega_{ab}^2
    -\frac{\omega_{ab}}{4z\sqrt{1+(az)^2}}\sin
\biggl(\,\frac{2\omega_{ab}z
\sinh^{-1}(az)}{az}\,\biggl)\,\biggr].
    \label{Accrr}
  \eea
The above result reduces to that of an inertially moving atom,
i.e., Eq.~(\ref{result form of rr}),  when $a$ goes to zero and it
vanishes if the boundary is approached. However, the most distinct
feature with the presence of a boundary is that the contribution
of radiation reaction now depends on the acceleration of the atom,
in sharp contrast to the unbounded Minkowski space where it has
been shown that for accelerated atoms on arbitrary stationary
trajectory, the contribution of radiation reaction is generally
not altered from its inertial value \cite{2Audretsch95}.

Adding up the two contributions, one finds the total rate of
change of the atomic excitation energy

\bea \left\langle {d H_A \/ d\tau} \right\rangle_{tot} &=& {\mu^2
\/ 2\pi}
    \Biggl[-  \sum_{\omega_a > \omega_b} \omega_{ab}^2 |\langle a
    | R_2^f (0) | b \rangle |^2 \;f(\omega_{ab},a,z)\;\left( 1 +{1\/ e^{{2\pi \/a}
\omega_{ab} }-1}
    \right) \nn\\
    &\qquad&\qquad + \sum_{\omega_a < \omega_b} \omega_{ab}^2
|\langle a
    | R_2^f (0)|b \rangle |^2 \;f(\omega_{ab},a,z)\;{1\/ e^{{2\pi \/a} |\omega_{ab}|
}-1} \;
    \Biggr] . \label{total}
    \eea
For an excited atom, only $\omega_{a}> \omega_{b}$ contributes.
One can see that the spontaneous emission is modified by the
appearance of the thermal term as compared to an inertial atom
near a reflecting boundary on one hand, and modified by the
appearance of $f(\omega_{ab},a,z)$ when compared to a uniformly
accelerated atom in an unbounded Minkowski space on the other.
However, for a ground-state atom, the delicate balance between the
vacuum fluctuations and radiation reaction no longer exists,
although both contributions of the vacuum fluctuations and
radiation are altered for accelerated atoms in the presence of the
boundary, as opposed to no change in the contribution of radiation
reaction in absence of boundaries. There is a positive
contribution from the $\omega_{a}< \omega_{b}$ term, therefore
transitions of ground-state atoms to excited states are allowed to
occur even in vacuum. The presence of the boundary modulates the
transition rate with the function $f(\omega_{ab},a,z)$ and makes
the rate a function of $z$, the atom distance from the boundary.
It is interesting to note that the spontaneous excitation rate of
accelerated atoms (or the Unruh effect) becomes smaller and
smaller as the atom is placed closer and closer to the boundary,
since $f(\omega_{ab},a,z)$ approaches 0 as $z\rightarrow 0$ for
any finite value of $a$.

Now we wan to evaluate the Einstein coefficient. In the present
case, we have two competing spontaneous processes, i.e., the
spontaneous excitation and de-excitation. Thus there are two
Einstein coefficients $A_\downarrow$ and $A_\uparrow$ which
describe the corresponding transition rates.
 Consider an ensemble of $N$ atoms. Let $N_1$ denote
the number of atoms in the ground state, $N_2$ the number in the
excited state. The rate equations are given by
 \beq {d N_2 \/ d \tau} = - {d N_1 \/ d \tau} = A_{\uparrow} N_1 -
    A_{\downarrow} N_2
    \eeq
with
 \beq \langle H_A \rangle = {1\/N} \left( - {1\/2} \omega_0
N_1 + {1\/2}
    \omega_0 N_2 \right).
 \eeq
The solution of the above equations is
 \beq
 \langle H_A (\tau)
\rangle = - {1\/2} \omega_0 + \omega_0
    { A_{\uparrow} \/ A_{\uparrow} + A_{\downarrow}} + \left(
\langle H_A (0)
     \rangle + {1\/2} \omega_0 - { A_{\uparrow} \/ A_{\uparrow}
     +A_{\downarrow}}\omega_0 \right) e^{- ( A_{\uparrow} +
A_{\downarrow} )
     \tau }\;.\label{FormalSol}
     \eeq
On the other hand,  we can simplify Eq. (\ref{total}) to obtain a
differential equation for $\langle H_A \rangle$
 \beq \left\langle
{d H_A (\tau) \/ d\tau} \right\rangle = -{\mu^2 \/ 4\pi}
    \omega_0 \left( {1\/4} \omega_0 + f(\omega_0,a,z)\left( {1\/2} + {1\/
e^{{2\pi \/a}
    \omega_0} -1} \right)\langle H_A (\tau) \rangle \right)\;,
\label{RateEQ}
 \eeq
 the solution of which gives the time evolution of the mean
atomic excitation energy
 \bea \langle H_A (\tau) \rangle &=& -
{1\/2} \omega_0 + { \omega_0 \/
    e^{{2\pi \/a} \omega_0} +1}
     + \left( \langle H_A (0) \rangle + {1\/2} \omega_0 -  {
\omega_0 \/
    e^{{2\pi \/a} \omega_0} +1}\right). \, \nn\\
    && \exp\left[ -{\mu^2\/4
\pi} \omega_0
    \left( {1\/2} + { 1 \/ e^{{2\pi \/a} \omega_0} -1} \right)
\;f(\omega_0,a,z)\tau \right].
    \label{RateSol}
    \eea
This indicates that the atom evolves with a modified decay
parameter towards the equilibrium value
 \beq \langle H_A  \rangle
= - {1\/2} \omega_0 + { \omega_0 \/
    e^{{2\pi \/a} \omega_0} +1},
 \eeq
revealing a thermal excitation with temperature $T=a/2\pi$ above
the ground state. The + 1 in the denominator of the second term
suggests that the atom obeys Fermi-Dirac statistics in thermal
equilibrium. This remarkable feature can be understood as a result
of the fermionic nature of a two-level system (for example, the
atomic raising and lowering operators obey the anticommutation
relation $\{R_+,R_-\} =1$).

 The Einstein coefficients $A_\downarrow$ and
$A_\uparrow$ for an accelerated atom near a reflecting plane
boundary readily follows from (\ref{RateSol}) and
(\ref{FormalSol}),
 \beq
A_{\downarrow} = {\mu^2 \/ 8 \pi} \omega_0 \left( 1 + {1\/ e^{{2
\pi
    \/a} \omega_0} -1} \right)f(\omega_0,a,z), \qquad A_{\uparrow} = {\mu^2 \/ 8
\pi}
    \omega_0 {\;f(\omega_0,a,z)\/ e^{{2 \pi \/a} \omega_0}-1}. \label{EinsteinCoe}
    \eeq
A comparison of the coefficient $A_\downarrow$ for spontaneous
emission from an accelerated atom near the plane boundary with the
corresponding Einstein coefficients for an inertial atom near a
plane boundary obtained in the last section and for an accelerated
atom in an unbounded Minkowski space found in
Ref.~\cite{Audretsch94} shows that rate of spontaneous emission is
enhanced by the thermal contribution as compared to the inertial
case with the presence of the boundary and is modulated by the
function $f(\omega_0,a,z)$ as compared to the accelerating case in
an unbounded flat space. Since $f(\omega_0,a,z)$ is an oscillating
function, the spontaneous emission rate can either be enhanced or
weakened as compared to the case of an accelerated atom without
the presence of the boundary, depending on the value of
$f(\omega_0,a,z)$. At the meantime, it is easy to see that
 the transition rate $A_\uparrow$ for the spontaneous
excitation is nonzero as long as $a\neq 0$ or $z\neq 0$ and it
vanishes as $a\to 0$ as expected.

\section{Conclusions}

 In conclusion,  assuming a dipole like interaction between the
atom and a scalar quantum field, we have studied the spontaneous
emission of a two-level atom in a space with a reflecting plane
boundary and examined both the contributions of vacuum
fluctuations and radiation reaction for both inertial motion and
uniform acceleration following the method developed in
Refs.~\cite{Dalibard82,Dalibard84,Audretsch94}.

In the case of an inertial atom,  our results show that  for
ground-state atoms, the contributions of vacuum fluctuations and
radiation reaction to the rate of change of the mean excitation
energy $\langle H_A (\tau)\rangle$ cancel exactly and the
cancellation ensures that there are no upward radiative
transitions in vacuum. For any initial excited state,
 the rate of change of atomic energy acquires equal
contributions from vacuum fluctuations and from radiation reaction
regardless of the distance from the plane boundary (refer to
Eq.(\ref{result form of vf}) and Eq.(\ref{result form of rr}) ).
Therefore, the presence of a plane boundary does not change the
delicate balance between the effects of vacuum fluctuations and
radiation reaction and these two different effects seem to be
equally important  both in the unbounded Minkowski space and the
space a plane boundary.  Although the corrections induced by the
presence of a boundary does not change the physical picture
qualitatively, it does quantitatively. At distances far from the
plane $(z\omega_{ab}\gg1)$, the corrections become negligible as
one would expect. It is interesting to note that close to the
plane$(z\omega_{ab}\ll1)$, the corrections becomes so large that
the total radiation rate of the atom diminishes to zero in a
oscillatory manner as the boundary is approached.  Finally, let us
note that the oscillatory behavior of the spontaneous radiation
rate of an excited atom near a reflecting boundary may offer a
possible opportunity for experimental tests for geometrical
(boundary) effects in flat spacetime.

In the case of a uniformly accelerated atom, both contributions of
the vacuum fluctuations and radiation reaction are altered by the
presence of a reflecting plane boundary and the delicate balance
between these two contributions existing in the case of inertial
ground-state atoms is disturbed, making possible the spontaneous
excitations from ground states.  There are some interesting
features to be noted as compared to the case without any boundary.
First,  the rate of change of the mean atomic excitation energy is
now a function of the distance to the boundary and it dies off in
an oscillatory way the boundary is approached, and second, the
contribution of radiation reaction is now dependent on the
acceleration of the atom, in sharp contrast to the unbounded
Minkowski space where it has been shown that for accelerated atoms
on arbitrary stationary trajectory, the contribution of radiation
reaction is generally not altered from its inertial value
\cite{2Audretsch95}.

\vskip 0.5cm

\begin{acknowledgments}
This work was supported in part  by the National Natural Science
Foundation of China  under Grant No. 10375023, the Program for
NCET (No. 04-0784), the Key Project of Chinese Ministry of
Education (No. 205110), the Key Project of Hunan Provincial
Education Department (No. 04A030) and the National Basic Research
Program of China under Grant No. 2003CB71630.

\end{acknowledgments}


\end{document}